\documentclass[12pt]{article}
\usepackage{epsfig}
\begin{document}
\title{\bf Spherically Symmetric Gravitational Collapse}

\author{M. Sharif \thanks{msharif@math.pu.edu.pk} and  Khadija
Iqbal\\
Department of Mathematics, University of the Punjab,\\
Quaid-e-Azam Campus, Lahore-54590, Pakistan.}

\date{}

\maketitle

\begin{abstract}
In this paper, we discuss gravitational collapse of spherically
symmetric spacetimes. We derive a general formalism by taking two
arbitrary spherically symmetric spacetimes with $g_{00}=1$. Israel's
junction conditions are used to develop this formalism. The formulae
for extrinsic curvature tensor are obtained. The general form of the
surface energy-momentum tensor depending on extrinsic curvature
tensor components is derived. This leads us to the surface energy
density and the tangential pressure. The formalism is applied to two
known spherically symmetric spacetimes. The results obtained show
the regions for the collapse and expansion of the shell.
\end{abstract}

{\bf Keywords:} Gravitational collapse.

\section{Introduction}

It has been an interesting problem to investigate whether the
spacetime singularities can be observed or not. Penrose [1]
suggested a cosmic censorship conjecture to investigate this
problem. This conjecture ensures that end state of a gravitational
collapse must be a black hole. There is no theoretical or
mathematical proof available to this conjecture. It would be
worthwhile to investigate this issue for complete understanding of
gravitational collapse or to obtain its correct form. This can be
pursued by taking the appropriate geometry of exterior and interior
regions and proper junction conditions which allow the matching of
these two regions.

Many people [2-4] have discussed gravitational collapse by taking
appropriate geometry of interior and exterior regions. Oppenheimer
and Synder [5] were the pioneer to study the gravitational collapse.
They worked on dust collapse by taking static Schwarzschild in the
exterior and Friedmann like solution in the interior spacetime. The
conclusion was that the end state of gravitational collapse leads to
the formation of a black hole. Misner and Sharp [6] worked on the
gravitational collapse by taking a static exterior and a perfect
fluid in the interior. Markovic and Shapiro [7] generalized the
pioneer's work in the presence of a positive cosmological constant.
Lake [8] extended this work by taking both positive and negative
cosmological constants. Cissoko et al. [9] studied explicitly the
effects of a positive cosmological constant on gravitational dust
collapse. He found that there are two physical horizons instead of
one and also discussed the effects of cosmological constant on
gravitational collapse. Debnath et al. [10] studied quasi-spherical
collapse with a positive cosmological constant. They also discussed
the apparent horizons and their physical significance. Sharif and
Ahmad [11,12] studied gravitational collapse of a perfect fluid with
a positive cosmological constant by taking static exterior and
non-static interior spacetimes. The same authors generalized this
work to five and higher dimensional spacetimes [13,14].

In 1966, Israel [15,16] presented junction conditions by taking
interior and exterior spacetimes. A lot of work on gravitational
collapse has been done by using these conditions [17-20]. Villas da
Rocha et al. [21] studied the self-similar gravitational collapse of
perfect fluid using Israel's method. He concludes that when collapse
has continuous self-similarity, black hole formation starts with
zero mass and when the collapse has no self-similarity, black hole
formation starts with a finite non-zero mass. By the same procedure,
Pereira and Wang [22,23] discussed the gravitational collapse of
cylindrical shells made of counter rotating dust particles.
Recently, Sharif and Ahmad [24] generalized this work to plane
symmetric spacetime. They derived general formulae for arbitrary
plane symmetric spacetimes and then applied this formalism to known
plane symmetric spacetimes. These studies have provided some
interesting results about gravitational collapse. However, no
attempt has been made for spherically symmetric spacetimes.

In this paper, we extend this work to spherical symmetric spacetimes
with matching conditions. First of all, we develop a formalism for
general spherical symmetric spacetimes and then the results are
found by taking particular examples. The scheme of this paper is as
follows: In section \textbf{2}, we provide the derivation of general
formalism by considering two arbitrary spherically symmetric
spacetimes with $g_{00}=1$. In sections \textbf{3} and \textbf{4},
we shall implement these formulae to discuss gravitational collapse
by taking two physical spherically symmetric spacetimes. The last
section will provide summary and discussion of the results obtained.

\section{General Formalism}

We consider a time-like $3D$ hypersurface $\Sigma$ which divides a
$4D$ spherically symmetric spacetime into two regions named as
interior and exterior spacetimes. For the interior region denoted
by $V^-$, we take the line element in the form as
\begin{equation}
ds^2_-=dt^2-X^-(t,r)dr^2-Y^-(t,r)(d\theta^2+\sin^2\theta d\phi^2),
\end{equation}
where $\{\chi^{-\mu}\} \equiv \{t,r,\theta,\phi\}$ are the usual
spherical polar coordinates. The line element for the exterior
region $V^+$ is given by
\begin{equation}
ds^2_+=dT^2-X^+(T,R)dR^2-Y^+(T,R)(d\theta^2+\sin^2\theta d\phi^2),
\end{equation}
where $\{\chi^{+\mu}\} \equiv \{T,R,\theta,\phi\}$ is another set
of spherical polar coordinates.

According to junction condition [15,16,25,26], we assume that the
interior and exterior spacetimes are the same on the hypersurface
$\Sigma$ which can be expressed as
\begin{equation}
(ds^2)=(ds^2_-)_\Sigma=(ds^2_+)_\Sigma.
\end{equation}
The equations of hypersurface $\Sigma$ in the coordinates
$\chi^{\pm\mu}$ are written as
\begin{equation}
f_-(r,t)=r-r_0(t)=0,
\end{equation}
\begin{equation}
f_+(R,T)=R-R_0(T)=0.
\end{equation}
Using Eqs.(4) and (5), the interior and exterior metrics reduce to
the following form respectively
\begin{equation}
(ds^2_-)_\Sigma=[1-X^-(t,r_0(t)) {r_0'}^2(t)]
 dt^2-Y^-(t,r_0(t))(d\theta^2+\sin^2\theta d\phi^2),
\end{equation}
\begin{equation}
(ds^2_+)_\Sigma=[1-X^+(T,R_0(T)) {R_0'}^2(T)]
dT^2-Y^+(T,R_0(T))(d\theta^2+\sin^2\theta d\phi^2),
\end{equation}
where prime denotes the ordinary differentiation with respect to
the indicated argument. The metric on the hypersurface is written
as

\begin{equation}
(ds^2)_\Sigma=\gamma_{ab}d\xi^ad\xi^b=d\tau^2-Y(\tau)(d\theta^2+\sin^2d\phi^2).
\end{equation}
Here $\{\xi^a\}=\{\tau,\theta,\phi\},~(a=0,2,3)$ are the
coordinates of the hypersurface $\Sigma$  and $\tau$ denotes the
proper time of the surface. Making use of Eqs.(6)-(8) into the
junction condition (3), it follows that
\begin{eqnarray}
d\tau=[1-X^-(t,r_o(t)) {r_0'}^2(t)]^{\frac{1}{2}}dt
=[1-X^+(T,R_0(T)){R_0'}^2(T)]^{\frac{1}{2}}dT,
\end{eqnarray}
\begin{equation}
Y(\tau)=Y^-(t,r_o(t))=Y^+(T,R_o(T)).
\end{equation}
From Eqs.(4) and (5), the outward unit normals $n^\pm_\mu$ to the
hypersurface $\Sigma$ in the coordinates $\chi^{\pm}$ can be
evaluated as
\begin{equation}
n^-_\mu=[{\frac{X^-}{[1-X^- {r_0'}^2(t)]}}]^{\frac{1}{2}} (-r_0'(t),
1,0, 0),
\end{equation}
\begin{equation}
n^+_\mu=[{\frac{X^+}{[1-X^+{R_0'}^2(T)]}}]^{\frac{1}{2}}
(-R_0'(T), 1, 0, 0).
\end{equation}
The extrinsic curvature tensor $K^\pm_{ab}$ on the hypersurface
$\Sigma$ is defined as [22,23]
\begin{equation}
K^\pm_{ab}=n^\pm_\sigma(\frac{\partial^2
x^\sigma_\pm}{\partial\varepsilon^a\partial\varepsilon^b}+\Gamma^\sigma_{\mu\nu}
\frac{\partial x^\mu_\pm}{\partial\varepsilon^a}\frac{\partial
x^\nu_\pm} {\partial\varepsilon^b}),\quad(\sigma,\mu,\nu=0,1,2,3)
,\quad(a,b=0,2,3).
\end{equation}
From this expression, the non-vanishing components of $K^+_{ab}$ for
exterior spacetime are given by
\begin{eqnarray}
K^+_{\tau\tau}&=&\frac{(X^+)^{\frac{1}{2}}}{[1-X^+{R_0'}^2(T)]^{\frac{3}{2}}}
[R_0''(T) + \frac{{X^+}_{,T}}{X^+} R_0'(T)\nonumber\\& +&
\frac{{X^+}_{,R}}{
2X^+} {R_0'}^2(T) - \frac{{X^+}_{,T}}{2} {R_0'}^3(T)],\nonumber\\
K^+_{\theta\theta}&=&\frac{1}{2}
[\frac{X^+}{[1-X^+{R_0'}^2(T)]}]^{\frac{1}{2}}
[-\frac{{Y^+}_{,R}}{X^+}-{Y^+}_{,T} {R_0'}(T)],\nonumber\\
K^+_{\phi\phi}&=&\sin^2\theta K^+_{\theta\theta}.
\end{eqnarray}
The non-vanishing components of $K^-_{ab}$ for interior spacetime
can be obtained from the above expressions by making the
replacement
\begin{equation}
X^+,~Y^+,~R_0(T),~T,~R~\rightarrow~X^-,~Y^-,~r_0(t),~t,~r.
\end{equation}
In terms of $K^\pm_{ab}$ and $\gamma_{ab}$, the surface
energy-momentum tensor is defined as [15,16,22,23]
\begin{equation}
S_{ab}=\frac{1}{\kappa}\{[K_{ab}]-\gamma_{ab}[K]\},
\end{equation}
where $\kappa$ is the coupling constant and
\begin{eqnarray}
[K_{ab}]=K^+_{ab}-K^-_{ab},\quad [K]=\gamma^{ab}[K_{ab}].
\end{eqnarray}
Using Eq.(14) and the corresponding expressions for $K^-_{ab}$, we
can write $S_{ab}$ in the form
\begin{equation}
S_{ab}={\rho}\omega_a\omega_b+{p}
(\theta_a\theta_b+\phi_a\phi_b),\quad(a,b=\tau,\theta,\phi),
\end{equation}
where $\rho$ is the surface energy density, $p$ is the tangential
pressure provided that they satisfy some energy conditions [27] and
$\omega_a,\theta_a,~\phi_a$ are unit vectors defined on the surface
given by
\begin{equation}
\omega_a=\delta^\tau_a,\quad
\theta_a=Y^\frac{1}{2}\delta^\theta_a,\quad
\phi_a=Y^\frac{1}{2}\sin\theta\delta^\phi_a.
\end{equation}
Here $\rho$ and $p$ are given by
\begin{eqnarray}
\rho=\frac{2}{\kappa Y}[K_{\theta\theta}],\quad
p=\frac{1}{\kappa}\{[K_{\tau\tau}]-\frac{[K_{\theta\theta}]}{Y}\}.
\end{eqnarray}

\section{Application}

In this section, we apply the general formalism derived in the
previous section by considering two known spherically symmetric
spacetimes to discuss the gravitational collapse. For the interior
region, we take Minkowski spacetime given by
\begin{equation}
ds^2_-=dt^2-dr^2-r^2(d\theta^2+\sin^2\theta d\phi^2).
\end{equation}
For the exterior region, the metric is taken as [28]
\begin{equation}
ds^2_+=dT^2-dR^2-a^2(T-R)(d\theta^2+\sin^2\theta d\phi^2),
\end{equation}
where $a$ is an arbitrary function of $T-R$. We re-write the above
solution by taking $\epsilon\equiv T-R$ for the sake of simplicity
as
\begin{equation}
ds^2_+=dT^2-dR^2-a^2(\epsilon)(d\theta^2+sin^2\theta d\phi^2).
\end{equation}
Using Eqs.(21) and (23), the junction conditions (9) and (10) take
the form
\begin{equation}
d\tau=[1-{R_0'}^2(T)]^{\frac{1}{2}}dT=[1-{r_0'}^2(t)]^{\frac{1}{2}}dt,
\end{equation}
and
\begin{equation}
r_0(t)=a(\epsilon_0),
\end{equation}
where $\epsilon_0=T-R_0(T)$. From Eqs.(24) and (25), we have
\begin{equation}
(\frac{dT}{dt})^2=\frac{1}{\Delta^2}\equiv\frac{1}{[(1-{R_0
'}^2(T))+{a'}^2(\epsilon_0)(1-{R_0 '}(T))^2]},
\end{equation}
where
\begin{equation}
{\Delta} =\sqrt{[(1-{R_0 '}^2(T))+{a'}^2(\epsilon_0)(1-{R_0
'}(T))^2]}.
\end{equation}
From Eqs.(25) and (26), we get
\begin{eqnarray}
r_0''(t)&=&\frac{a ''(\epsilon_0)(1-{R_0 '}^2(T))^2-a
'(\epsilon_0)R_0 ''(T)}
{[(1-{R_0 '}^2(T))+{a'}^2(\epsilon_0)(1-{R_0'}(T))^2]}\nonumber\\
&-&\frac{1}{[(1-{R_0
'}^2(T))+{a'}^2(\epsilon_0)(1-{R_0'}(T))^2]^\frac{3}{2}}[{a '}
^2(\epsilon_0)a''(\epsilon_0)(1-R_0
'(T))^4\nonumber\\
&-&a'(\epsilon_0)R_0'(T)R_0''(T)(1-R_0'(T))- {a
'}^3(\epsilon_0)R_0 ''(T) (1-R_0 '(T))^2].\nonumber\\
\end{eqnarray}
Making use of Eq.(14) and the corresponding expression for
$K^-_{ab}$ into Eq.(20) and considering Eqs.(25)-(28), the
expressions for $\rho$ and $p$ take the form
\begin{eqnarray}
\rho&=&\frac{2}{\kappa a(\epsilon_0)(1-{R_0 '}
^2(T))^\frac{1}{2}}[a'(\epsilon_0)((1-R_0 '(T))+\Delta],\\
p&=&\frac{1}{\kappa a(\epsilon_0)\Delta(1-R_0
'(T))^\frac{3}{2}}[a(\epsilon_0)\Delta R_0
''(T)-a(\epsilon_0)a''(\epsilon_0)(1-{R_0
'}^2(T))\nonumber\\&&(1-R_0 '(T))^2+a(\epsilon_0)a'(\epsilon_0)
R_0 ''(T)(1-{R_0 '}^2(T))-a(\epsilon_0)
a'(\epsilon_0)\nonumber\\&&R_0 '(T)R_0 ''(T)(1-R_0
'(T))-2a'(\epsilon_0)\Delta (1-{R_0 '}^2(T))((1-R_0
'(T))\nonumber\\&&+\Delta)-2\Delta^2(1-{R_0 '}^2(T))].
\end{eqnarray}
To see the minimum effects of shell on the collapse, we take $p=0$
in Eq.(30), it follows that
\begin{eqnarray}
&&R_0''(T)=(1-{R_0
'}^2(T))\nonumber\\&&\times\frac{[a(\epsilon_0)a''(\epsilon_0)(1-R_0
'(T))^2+2a'(\epsilon_0)\Delta (1-R_0
'(T))+2\Delta^2]}{[a(\epsilon_0)\Delta-a(\epsilon_0)
a'(\epsilon_0)R_0 '(T)(1-R_0
'(T))+a(\epsilon_0)a'(\epsilon_0)(1-{R_0 '}^2(T))]}.\nonumber\\
\end{eqnarray}

To investigate the spherically symmetric gravitational collapse, we
need to solve the above equation for $R_0(T)$. However, it is much
more complicated to solve this in general to get any insight.
Therefore, we consider a particular case in which
\begin{eqnarray}
R_0''(T)=(1-{R_0 '}^2(T)\beta,
\end{eqnarray}
where $\beta$ is assumed to be an arbitrary constant given by
\begin{eqnarray}
\beta=\frac{[a(\epsilon_0)a'(\epsilon_0)'(1-R_0
'(T))^2+2a'(\epsilon_0)\Delta (1-R_0
'(T))+2\Delta^2]}{[a(\epsilon_0)\Delta-a(\epsilon_0)
a'(\epsilon_0)R_0 '(T)(1-R_0
'(T))+a(\epsilon_0)a'(\epsilon_0)(1-{R_0 '}^2(T))]}.
\end{eqnarray}
Now integrating Eq.(32), we obtain
\begin{eqnarray}\label{2.2.24}
R_0'(T)=\frac{e^{2\beta T}+e^{2c_1}}{e^{2\beta T}-e^{2c_1}},
\end{eqnarray}
where $c_1$ is constant of integration. Here, we consider the
following two cases:
\begin{eqnarray}\label{2.2.25}
(I)~~\beta >0\quad~~~~~~~~~~~(II)~~\beta <0.
\end{eqnarray}

\subsection{Case I}

In this case, we take $\beta=1$ and $c_1=0$ for simplicity, then
Eq.(\ref{2.2.24}) takes the form
\begin{equation}\label{2.2.26}
R_0'(T)=\frac{e^{2T}+1}{e^{2T}-1}.
\end{equation}
Integration of the above equation yields
\begin{eqnarray}\label{2.2.27}
R_0(T)=-T+c_2+\ln(-1+e^{2T}),
\end{eqnarray}
where $c_2$ is also a constant of integration.  For $c_2=0$,
Eqs.(\ref{2.2.26}) and (\ref{2.2.27}) yield respectively
\begin{eqnarray}\label{2.2.28}
R_{0}(T)&=&\left\{ \begin{array}{lll}
0.04132&,&T=0.5  \\
+\infty&,&T=+\infty,
\end{array}\right.
\\
\label{2.2.29} R_{0}'(T)&=&\left \{ \begin{array}{lll}
2.16359&,&T=0.5  \\
1&,&T=+\infty.
\end{array} \right.
\end{eqnarray}
\begin{figure}
\center\epsfig{file=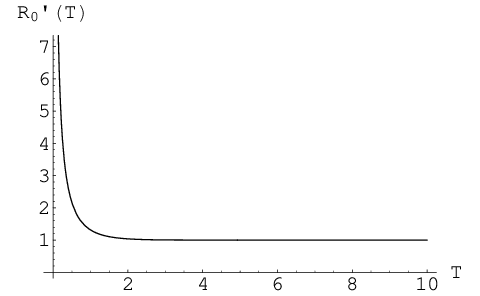,width=0.5\linewidth} \caption{
velocity-time graph for $c_1=0$. }
\end{figure}
\begin{figure}
\center\epsfig{file=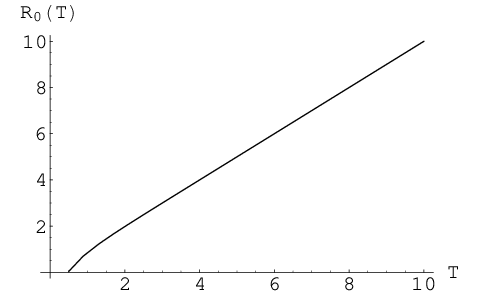,width=0.47\linewidth} \caption{
displacement-time graph for $c_2=0$. }
\end{figure}
This case gives expanding process. It can be seen from figures 1 and
2 that radial velocity is positive and radius increases from some
finite value to infinity. The expansion starts from some finite
value at some finite time with finite positive velocity and ends at
$T=+\infty$, where radial velocity is unity. It is also observed
that by considering different values of $c_2$ we get the same
expanding process. We can take different values of $c_2$. The
difference for possibilities of $c_2$ is that by increasing the
values of $c_2$, the time interval for the expanding process
increases.\\

\subsection{Case II}

Here we take $\beta =-1$ and $ c_1=0$ for simplicity, then
Eq.(\ref{2.2.24}) implies that
\begin{equation}\label{2.2.42}
R_0'(T)=\frac{e^{-2T}+1}{e^{-2T}-1}.
\end{equation}
Integration of this equation gives
\begin{eqnarray}\label{2.2.43}
R_0(T)=T+c_3-\ln(-1+e^{2T}),
\end{eqnarray}
where $c_3$ is a constant of integration. For $c_3=0$,
Eqs.(\ref{2.2.42}) and (\ref{2.2.43}) give
\begin{eqnarray}\label{2.2.44}
R_{0}(T)&=&\left \{\begin{array}{lll}
+\infty&,&T=0 \\
0.000026&,&T=0.4812,
\end{array} \right.
\\
\label{2.2.45} R'_{0}(T)&=&\left \{\begin{array}{lll}
-\infty&,&T=0 \\
-1.45678&,&T=0.4812.
\end{array} \right.
\end{eqnarray}
\begin{figure}
\center\epsfig{file=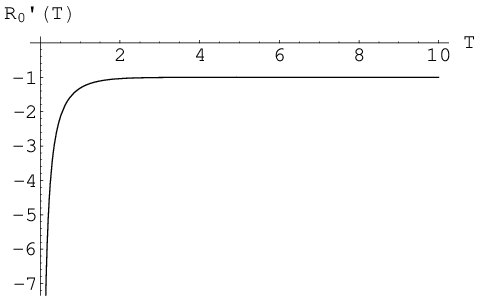,width=0.47\linewidth} \caption{
velocity-time graph for $c_1=0$.}
\end{figure}
\begin{figure}
\center\epsfig{file=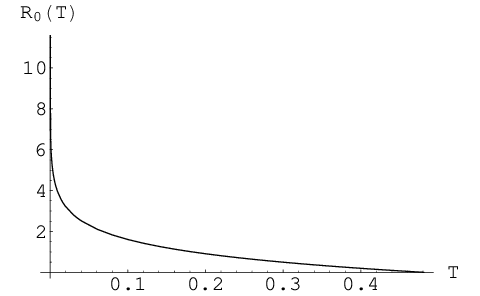,width=0.47\linewidth} \caption{
displacement-time graph for $c_3=0$.}
\end{figure}
This case represents a collapsing process. It can be seen from
figures 3 and 4 that radial velocity is negative and radius
positively decreases from infinity to some finite value. The shell
starts collapsing from some finite time with infinite negative
velocity and ends with some finite radial velocity at some finite
time. It is also found that different values of $c_3$ shows the same
collapsing process. The difference between possibilities of $c_3$ is
that by increasing the values of $c_3$, the time interval for
collapsing process increases.\

\section{Summary and Discussion}

This paper is devoted to the study of spherically symmetric
gravitational collapse using Israel's method. We have taken two
arbitrary spherically symmetric spacetimes with $g_{00}=1$. We have
developed a general formalism for extrinsic curvature and surface
energy-momentum tensor in terms of the metric coefficients and their
first derivatives by using Israel's junction condition. As an
application of this formalism, we have taken Minkowski spacetime as
an interior region whereas a non-static spacetime is taken as an
exterior region. The surface energy density and the tangential
pressure have been found. In order to see the minimum effects of the
shell on the collapse, we have taken the tangential pressure $p=0$.

There arise two main cases, i.e., \textbf{Case I} and \textbf{Case
II}. In the \textbf{Case I}, we have taken an arbitrary constant
$\beta>0$ and for simplicity its value as unity. Here, we get an
expanding process with positive radial velocity and positively
increasing radius. The expansion starts from some finite value at
some finite time with finite positive velocity. The expansion ends
at $T=+\infty$, where radial velocity is unity. It is also observed
that time interval for radial velocity decreases with the increase
of the constant used in Eq.(\ref{2.2.24}). In the nutshell, we can
say that we get an expanding sphere for all values of arbitrary
constant $\beta>0$ and the integration constants.

For the \textbf{Case II}, we have taken $\beta<0$ and in particular
$\beta=-1$ for simplicity. This case represents a collapsing process
with negative radial velocity and radius decreases positively from
infinity to some finite value. The shell starts collapsing from some
finite time with infinite negative velocity and ends with some
finite radial velocity at some finite time. Results also show that
by increasing the value of the constant used in Eq.(\ref{2.2.24}),
the time interval for the radial velocity also increases. Thus it
turns out that the case II represents a collapsing sphere for all
values of arbitrary and integration constants. It is interesting to
mention here that in the cases where collapse occurs, the energy
density remains finite. Thus the collapse does not end as a
singularity. It would be interesting if this formalism is developed
without taking $g_{00}=1$ so that more general examples could be
discussed.

\vspace{0.5cm}
\begin{description}
\item  {\bf Acknowledgment}
\end{description}

We would like to thank Mr. Zahid Ahmad for the fruitful discussions
on the subject.

\vspace{2cm}
%\newpage

{\bf \large References}

\begin{description}

\item{[1]} Penrose, R.: Riv. Nuovo Cimento \textbf{1}(1969)252.

\item{[2]} Ghosh, S.G. and Deshkar, D.W.: Int. J. Mod. Phys. \textbf{D12}(2003)317.

\item{[3]} Ghosh, S.G. and Deshkar, D.W.: Gravitation and Cosmology \textbf{6}(2000)1.

\item{[4]} Debnath, U., Nath, S. and Chakraborty, S.: Mon. Not. R.
Astron. Soc. \textbf{369}(2006)1961.

\item{[5]} Oppenheimer, J.R. and Synder, H.: Phys. Rev.
\textbf{56}(1939)455.

\item{[6]} Misner, C.W. and Sharp, D.: Phys. Rev. \textbf{136}(1964)B571.

\item{[7]} Markovic, D. and Shapiro, S.L.: Phys. Rev. \textbf{D61}(2000)084029.

\item{[8]} Lake, K.: Phys. Rev. \textbf{D62}(2000)027301.

\item{[9]} Cissoko, M., Fabris, J.C., Gariel, J., Denmat, G.L. and
Santos, N.O.: arXiv:gr-qc/9809057.

\item{[10]} Debnath, U., Nath, S. and Chakraborty, S.: Gen.
Relativ. Grav. \textbf{D37}(2005)215.

\item{[11]} Sharif, M. and Ahmad, Z.: Mod. Phys. Lett.
\textbf{A22}(2007)1493.

\item{[12]} Sharif, M. and Ahmad, Z.: Mod. Phys. Lett.
\textbf{A22}(2007)2947.

\item{[13]} Sharif, M. and Ahmad, Z.: J. Korean. Phys. Soc.
\textbf{52}(2008)980.

\item{[14]} Sharif, M. and Ahmad, Z.: Acta Physica Polonica B \textbf{39}(2008)1337.

\item{[15]} Israel, W.: Nuovo Cimento \textbf{B44}(1966)1.

\item{[16]} Israel, W.: Nuovo Cimento \textbf{B48}(1967)463(E).

\item{[17]} Lemos, J.P.S.: Phys. Rev. \textbf{D57}(1998)4600.

\item{[18]} Cai, R.G. and Wang, A.: Phys. Rev. \textbf{D73}(2006)063005.

\item{[19]} Herrera, L. and Santos, N.O.: Phys. Rev. \textbf{D70}(2004)084004.

\item{[20]} Herrera, L. and Santos, N.O.: Class. Quantum Grav.\textbf{22}(2005)2407.

\item{[21]} Villas de Rocha, J.F., Wang, A. and Santos, N.O.: Phys. Lett.
\textbf{A255}(1999)213.

\item{[22]} Pereira, P.R.C.T. and Wang, A.: Phys. Rev. \textbf{D62}(2000)124001.

\item{[23]} Pereira, P.R.C.T. and Wang, A.: Phys. Rev. \textbf{D67}(2003)129902(E).

\item{[24]} Sharif, M and Ahmad, Z.: Int. J. Mod. Phys. \textbf{A23}(2008)181.

\item{[25]} Santos, N.O.: Phys. Lett. \textbf{A106}(1984)296.

\item{[26]} Santos, N.O.: Mon. Not. R. Astr. Soc. \textbf{216}(1985)403.

\item{[27]} Hawking, S.W. and Ellis, G.F.R: {\it The Large Scale Structure of Spacetime}
(Cambridge University Press, Cambridge, 1973).

\item{[28]} Qadir, A.: J. Math. Phys. \textbf{33}(1992)6.

\end{description}
\end{document}